\documentclass[aps,prl,twocolumn]{revtex4-1}
\usepackage{hyperref}
\hypersetup{colorlinks=true, citecolor=blue, urlcolor=blue, linkcolor=blue}

\pdfoutput=1
\usepackage[utf8]{inputenc}
\usepackage[T1]{fontenc}
\usepackage{calc}
\usepackage{graphics,graphicx,amsfonts,amsmath,amsbsy,amssymb,color}
\usepackage{bm}
\usepackage{subfigure}
\usepackage{threeparttable}

\begin{document}

\title{Structural and electronic properties of solid molecular hydrogen from many-electron theories}

\author{Ke Liao}
\thanks{These two authors contributed equally}
\affiliation{
Max Planck Institute for Solid State Research, Heisenbergstrasse 1, 70569 Stuttgart, Germany
}

\author{Tong Shen}
\thanks{These two authors contributed equally}
\affiliation{State Key Laboratory for Artificial Microstructure and Mesoscopic Physics, Frontier Science Center for Nano-optoelectronics and School of Physics, Peking University, 100871 Beijing, China}

\author{Xin-Zheng Li}
\affiliation{State Key Laboratory for Artificial Microstructure and Mesoscopic Physics, Frontier Science Center for Nano-optoelectronics and School of Physics, Peking University, 100871 Beijing, China}
\affiliation{Collaborative Innovation Center of Quantum Matter, Peking University, 100871 Beijing, China}

\author{Ali Alavi}
\affiliation{
Max Planck Institute for Solid State Research, Heisenbergstrasse 1, 70569 Stuttgart, Germany
}

\affiliation{
Department of Chemistry, University of Cambridge, Lensfield Road, Cambridge CB2 1EW, United Kingdom
}

\author{Andreas Gr\"uneis}
\email{andreas.grueneis@tuwien.ac.at}
\affiliation{
  Institute for Theoretical Physics, TU Wien,\\
  Wiedner Hauptstraße 8-10/136, 1040 Vienna, Austria
}

\begin{abstract}

We study the structural and electronic properties of
phase III of solid hydrogen using accurate many-electron theories and
compare to state-of-the-art experimental findings.
The atomic structures of phase III modelled by C2/c-24 crystals are fully optimized
on the level of second-order perturbation theory, demonstrating that previously employed structures optimized on the level
of approximate density functionals exhibit errors in the H$_2$ bond lengths
that cause significant discrepancies in the computed
quasi particle band gaps and vibrational frequencies compared to experiment.
Using the newly optimized atomic structures,
we study the band gap closure and change in vibrational frequencies as a function of pressure.
Our findings are in good agreement with recent experimental observations and may prove useful
in resolving long-standing discrepancies between experimental estimates of metallization pressures
possibly caused by disagreeing pressure calibrations.



\end{abstract}

\maketitle


The seminal work of Wigner and Huntington -- that first predicted a metallization
of hydrogen~\cite{wigner1935} in 1935 at a pressure of about 25~GPa --
has sparked a continuous interest in the pressure-temperature phase diagram of hydrogen.
However, state-of-the-art experiments~\cite{hemley1989,narayana1998,loubeyre2002} have not been
able to conclusively detect metallic behaviour with the exception of some recent
experimental studies~\cite{dias2017,dias2017a,loubeyre2020} that are still under debate~\cite{geng2017,silvera2019}.
Until today, one of the most reliable experimental estimates for the metallization pressure
range is approximately 
425~GPa-450~GPa~\cite{loubeyre2020}. The lower value was obtained by the discontinuous 
pressure evolution in the infrared absorption, assuming a structural phase transition to
the atomic structure, whereas the higher value was obtained by extrapolation of the band gap,
assuming hydrogen remains in phase III.
Determining the metallization pressure accurately is extremely challenging.
This is partly reflected by the disagreement of the measured
H$_2$ vibron frequency peaks as a function of the pressure, which is crucial for
pressure calibration in many experiments~\cite{silvera2019}.
In addition to the electronic structure, questions concerning the atomic structure are also difficult to address.
Using X-ray scattering to determine the crystal structure experimentally is
hampered by the low scattering cross section of hydrogen.
Depending on pressure and temperature, hydrogen has been predicted to condense in
different orientationally ordered molecular crystals~\cite{hemley1988,lorenzana1989,mao1994,natoli1995,goncharenko2005,pickard2007,howie2012} or
(liquid) metallic~\cite{wigner1935,hemley1989,weir1996,johnson2000,eremets2011,chen2013,knudson2015,zaghoo2016,dias2017} phases.

Accurate theoretical predictions of the equilibrium phase boundaries and other properties of
high pressure hydrogen require an appropriate treatment of
quantum nuclear and many-electron correlation effects~\cite{li2013,drummond2015,azadi2013,mcminis2015,morales2013,morales2013a},
which can only be achieved using state-of-the-art \emph{ab initio} methods.
Hitherto, most \emph{ab initio} studies of solid hydrogen are based either on density functional
theory (DFT)~\cite{chacham1991,kohanoff1997,johnson2000,pickard2007} or quantum Monte Carlo
calculations~\cite{ceperley1987,azadi2013,clay2014,azadi2014,drummond2015,azadi2017,mcminis2015}.
DFT employing approximate exchange and correlation (XC) energy functionals can be applied to
compute infrared and Raman spectra as well as equilibrium phase boundaries, facilitating a direct
comparison between theory and experiment~\cite{hemley1988,mao1994,zha2012,lorenzana1989,hanfland1993,howie2012,monserrat2016,zhang2018}.
However, different parameterisations of the XC functional in DFT yield
inconsistent predictions~\cite{azadi2017c,clay2014,mcminis2015}.
Diffusion Monte Carlo (DMC) produces more reliable pressure temperature phase
diagrams~\cite{azadi2013,azadi2014,drummond2015,azadi2017,mcminis2015}.
Furthermore DMC can also be used to compute quasi particle gaps
including nuclear quantum effects~\cite{gorelov2020}.
Recently we have shown that coupled cluster singles and doubles (CCSD) theory
predicts static lattice enthalpies of solid hydrogen phases with high accuracy and computational efficiency~\cite{liao2019}.
CCSD results for the most stable model phases including phase II and III
are in good agreement with those obtained using diffusion Monte Carlo.
However, these studies are based on structures optimised using approximate XC functionals,
causing uncontrollable errors when comparing computed transition pressures or band gaps to experiment.
Here, we employ accurate many-electron theories to predict the atomic structure
of crystalline molecular hydrogen phases and related properties, enabling a more rigorous study of
band gaps and vibrational frequencies.

{\bf Methods.}
We optimise the atomic structure of model phase III using nuclear gradients calculated on the level of
second-order M\o ller-Plesset (MP2) perturbation theory and a plane wave basis set~\cite{supp}. 
We note there are some earlier implementions of MP2 forces in periodic solids using Gaussian basis set~\cite{del2015,weigend1997,rybkin2016}.
All periodic calculations have been performed using
the Vienna \emph{ab initio} simulation package (\texttt{VASP})~\cite{kresse1996}
in the framework of the projector augmented wave method~\cite{blochl1994},
interfaced to our coupled cluster code~\cite{hummel_2017} that employs an
automated tensor contraction framework (\texttt{CTF})~\cite{solomonik2014}.
We use Hartree--Fock orbitals in all post Hartree--Fock methods~\cite{gruneis2011}.
Computational details are discussed in Ref.~\cite{supp}.
Although MP2 theory can be considered a low-order approximation to CCSD theory,
it predicts lattice constants for a wide range of solids with higher accuracy than DFT-PBE when
compared to experiment~\cite{gruneis2010}.
Due to the many-electron nature of the employed Ansatz, CCSD theory is exact for two-electron systems.
The coupling between electron pairs is, however, approximated by truncating the many-body perturbation
expansion in a computationally efficient manner and performing
a resummation to infinite order of certain contributions only.

Phase III is modelled by C2/c-24 crystals~\cite{pickard2007}
initially predicted by {\em ab initio} simulations and
random structure searches~\cite{pickard2007,drummond2015}.
The structure is labelled by its symmetry followed
by the number of atoms in the primitive cell. C2/c-24 consists of layered hydrogen molecules whose
bonds lie within the plane of the layer, forming a distorted hexagonal shape.
We note that previous DMC studies employed structures that have been optimized using a
range of approximate density functionals,
indicating that an appropriate choice is crucial~\cite{mcminis2015}.
In this work we employ supercells containing up to 96 atoms for the relaxation of the atomic positions.
The convergence with respect to computational parameters
such as number of virtual orbitals, $k$-meshes for the Hartree--Fock energy contribution
and energy cut offs for the employed plane wave
basis set have been checked carefully and are summarized in the supplemental materials~\cite{supp}.

Beyond the static lattice model, the $T$-dependent band gap renormalization of the 
single particle excitation energy due to electron-phonon interactions (EPIs) was also studied, using a
dynamical extension of the static EPIs theory originally proposed by Heine, Allen, and Cardona (HAC)~\cite{allen1976,allen1983}. 
The quasiparticle approximation (QPA) was used to correct the DFT-PBE eigenvalues based on the EPIs self-energies.
These calculations are performed 
using \texttt{QUANTUM ESPRESSO (QE)}~\cite{giannozzi2009} and \texttt{YAMBO}~\cite{marini2009,sangalli2019}. 
The excitonic effects were obtained by solving the Bethe-Salpeter equation (BSE), as implemented
in \texttt{VASP}. The EPIs and the excitonic effects are calculated using DFT-PBE optimized 
primitive cell structure.

{\bf Results.}
We fully relax the internal degrees of freedom
of DFT-PBE structures by minimizing the atomic forces
computed on the level of MP2 theory, while keeping the lattice vectors fixed and
maintaining the space group symmetry.
The MP2 structures are published in the supplemental information alongside additional results,
demonstrating that further effects resulting from the relaxation of the lattice vectors can be disregarded~\cite{supp}.
For the purpose of the following discussion we will focus on the shortest hydrogen bond length
in these structures, which represents the most striking difference between MP2 and DFT-PBE results.
At a pressure of 250~GPa, the shortest hydrogen molecule bond length in 
the DFT-PBE structures for phase III
is 0.75~\AA\ , whereas MP2 theory predicts 0.72~\AA.
Similar findings apply to the structures at other pressures.
In passing we note that the shortest hydrogen molecule bond length obtained using the
vdW-DF functional~\cite{lee2010} is 0.72~\AA, which is fortuitously close to
our MP2 findings and agrees with findings
reported in Ref.~\cite{mcminis2015}.
However, it is important to assess the reliability of these newly
optimised structures further by comparing to CCSD results.
\begin{figure}[t]
  \begin{center}
\includegraphics[width=0.99\linewidth]{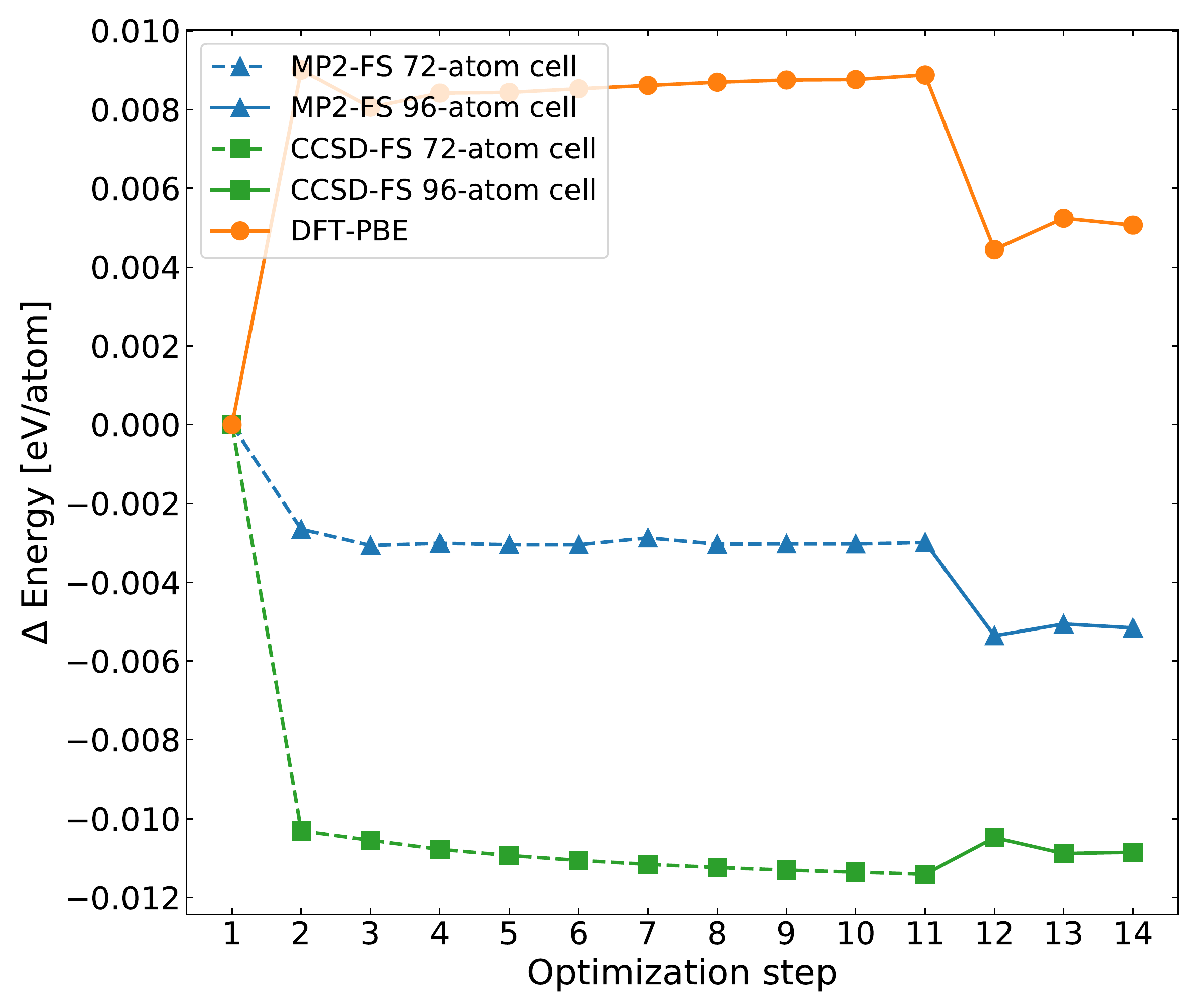}
  \end{center}
\caption{
  The energy changes during the structural relaxation as a function of the optimization steps.
  This example at the DFT pressure of 250 GPa shows that
  the MP2 and CCSD total energies per atom
  are lowered in a similar
  fashion and provides evidence that the optimized MP2 structures are close to the
  CCSD structures.
  The MP2 and CCSD energies are corrected by finite-size
  corrections~\cite{gruber2018} and are labelled by MP2-FS
  and CCSD-FS, respectively,
}
\label{fig:structure}
\end{figure}
Fig.~\ref{fig:structure} illustrates that the total MP2 energy per atom of phase III at a volume
of 1.57~\AA$^3$/atom  (corresponding to a DFT-PBE pressure of 250~GPa) is lowered by
about 5~meV/atom during the structural relaxation. The initial 11 steps of the relaxation were carried out using a 72 atom supercell
only, whereas all further optimization steps have been performed using a 96 atom supercell, indicating that finite size
effects become negligible.
The shortest bond length is only changed
by about 0.01~{\AA}  betweeen the 11th and the final step.  
After 14 steps the remaining forces on the atoms are smaller than 0.05~eV/\AA.
Fig.~\ref{fig:structure} also depicts  that the  CCSD energy is lowered in total by 11~meV/atom during the full
MP2 relaxation trajectory, which is similar to the change in MP2 theory.
The latter observation is important because it demonstrates that MP2 and CCSD equilibrium structures are expected to
deviate only slightly. This justifies the main assumption of the present work which states that MP2
structures for phase III are very accurate.
To further substantiate this claim, we note that 
MP2 theory  predicts lattice constants for a wide range of solids with significantly higher accuracy than DFT-PBE when
compared to experiment~\cite{gruneis2010}.

\begin{figure}[t]
\begin{center}
\includegraphics[width=0.99\linewidth]{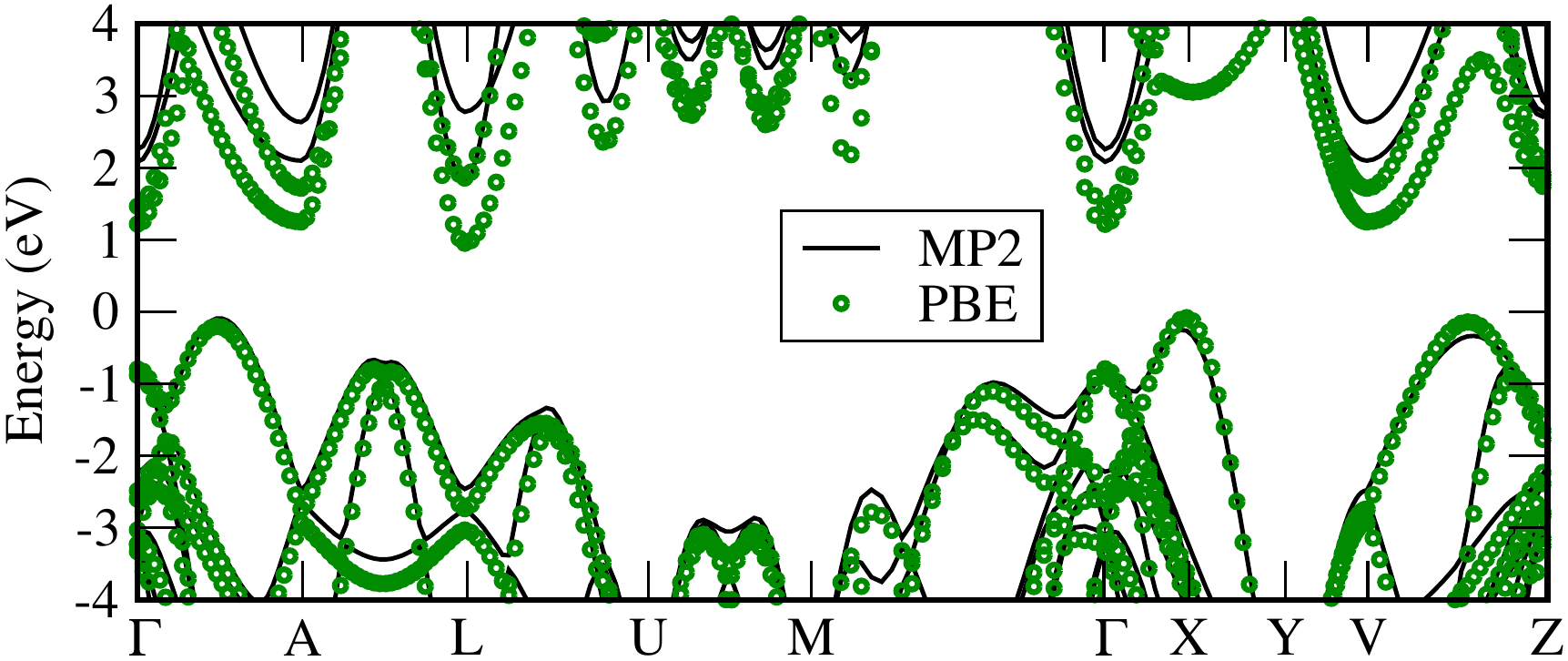}
\end{center}
\caption{\label{fig:bandstructure}
Electronic band structure of model phase III (C2/c-24) obtained using DFT-PBE.
Black (green) lines correspond to MP2 (DFT-PBE) equilibrium geometries at the pressure
of 250~GPa.
}
\end{figure}

As a first demonstration for the far-reaching consequences of the structural changes, we discuss
its impact on the quasi particle band gap of model phase III (C2/c-24).
Fig.~\ref{fig:bandstructure} depicts the electronic band structure for phase III at a pressure of 250~GPa employing the
atomic structures optimised using DFT-PBE and MP2 theory. The Kohn-Sham band structures are computed using the PBE functional,
exhibiting an indirect band gap with the valence band maximum at $X$ and the
conduction band minimum at $L$. The direct gap is located at $\Gamma$.
The direct and indirect PBE band gaps for the MP2 structure are 2.97~eV and 1.9~eV, respectively.
However, due to the reduced hydrogen bond length, the direct and indirect band gaps are about 1~eV larger
in the MP2 structure compared to the DFT-PBE structure.
We note that this increase in the band gap persists for the more accurate quasi particle band gaps computed on the
level of the $G_0W_0$ approximation.
We stress that due to the strong dependence of the electronic gap on the pressure,
an underestimation of the band gap by 1~eV results in a decrease in the predicted
metallization pressure by more than 50~GPa.
We note that the previously employed vdW-DF structures in
Refs.~\cite{mcminis2015,gorelov2020} yield band gaps that agree with
our findings obtained using the MP2 structures to within about 0.1~eV.
The direct and indirected PBE band gaps computed using the vdW-DF structures are 2.88~eV and 1.74~eV, respectively.

We now turn to the comparison between computed $G_0W_0$ band gaps and experimental findings.
As shown in Ref.~\cite{gorelov2020}, the inclusion of zero point vibrational effects to the quasi particle gaps is crucial.
At 0 K, this is termed as zero-point renormalization (ZPR).
At finite $T$s, $T$-dependent band gap renormalization also exists, originating from the Fan 
and Debye-Waller terms as described in the dynamical HAC theory. More details can be found in 
Ref.~\cite{sangalli2019, shen2020}.
Unfortunately a seamless inclusion of the electron-phonon coupling contributions to the band
gap on the level of MP2 theory would be computationally too expensive at the moment.
Therefore we estimate these renormalizations using DFT-PBE
phonons and include them in the $G_0W_0$ quasi particle band gaps~\cite{supp}.
Our calculations yield a ZPR of the direct and indirect gap of about -1~eV, which is
by coincidence a similar magnitude as the band gap increase caused by structural relaxation
but significantly smaller in magnitude than the -2~eV ZPR reported previously~\cite{gorelov2020}.
\begin{figure}[t]
\begin{center}
\includegraphics[width=0.99\linewidth]{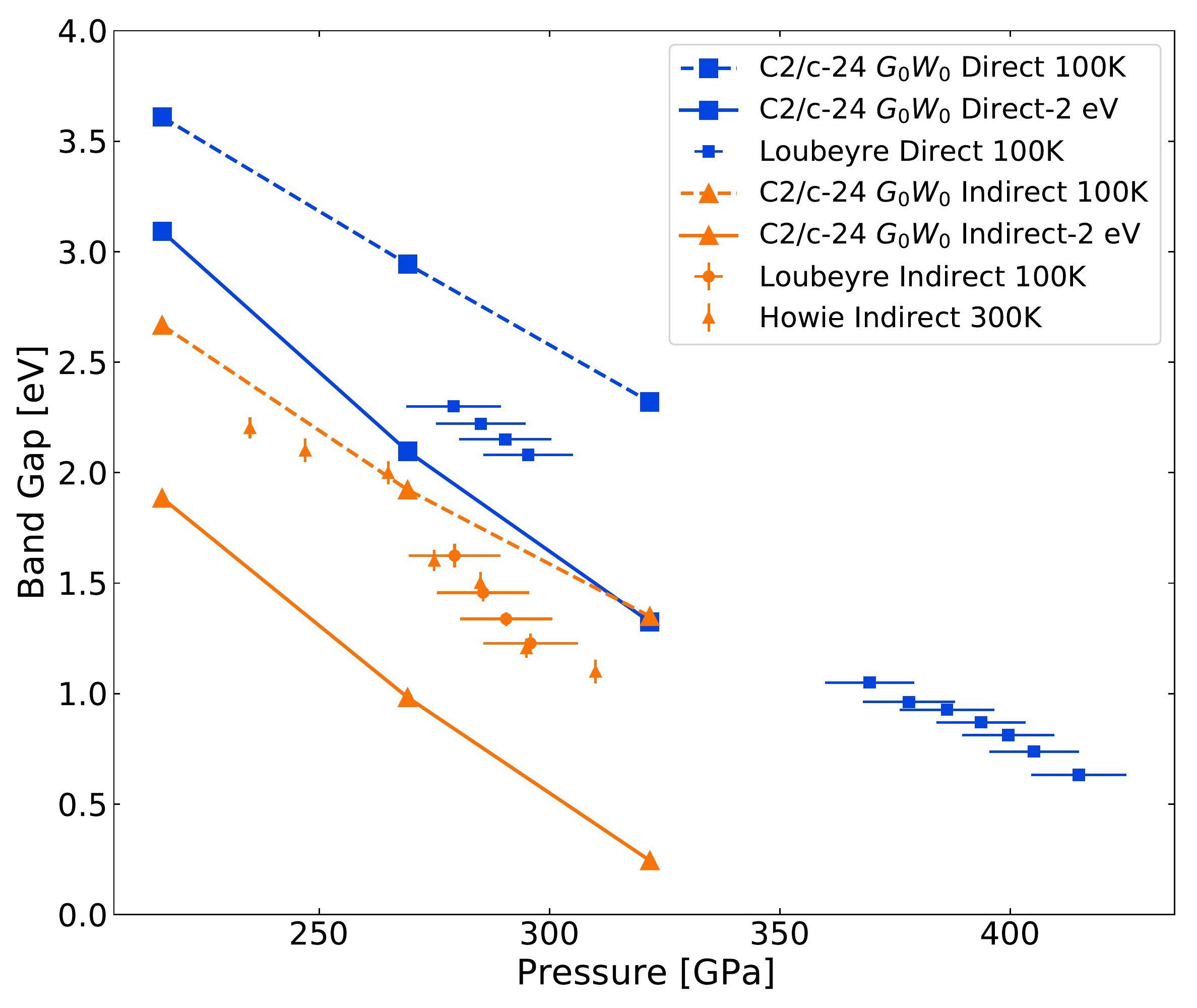}
\end{center}
\caption{\label{fig:gapvspressure}
Direct and indirect $G_0W_0$ band gaps including EPI contributions from this work (dashed lines) and $\approx -2$~eV EPI contributions
from Ref.~\cite{gorelov2020} (full lines). The direct $G_0W_0$ band gaps include $\approx$-0.12~eV
exciton binding energy.
The experimental estimates have been taken from Ref.~\cite{loubeyre2002, loubeyre2020,
howie2012, gorelov2020}.
}
\end{figure}
The difference of the $T$-dependent indirect band gap renormalizations in Ref.~\cite{gorelov2020} between 200 K and 300 K is about 0.2 eV,
which is an order of magnitude larger than our estimate of 0.02 eV~\cite{supp}. 
The difference in the experimental indirect band gaps between 
100 K~\cite{loubeyre2002} and 300 K~\cite{howie2012} is about 0.02~eV, which agrees much 
better with our result.
The computed $G_0W_0$ gaps with EPIs are depicted in Fig.~\ref{fig:gapvspressure} for a
range of pressures alongside experimental findings\cite{loubeyre2002,loubeyre2020,howie2012,gorelov2020}.
We note that the direct $G_0W_0$ gaps includes $\approx$-0.12~eV exciton
binding energy~\cite{supp} in order to enable a direct
comparison to the optical measurements from Ref.~\cite{loubeyre2002,loubeyre2020}.
Furthermore we plot the $G_0W_0$ gaps with respect to the CCSD pressures computed
from the enthalpy versus volume curves, enabling an accurate and direct
comparison to experimental findings.
Compared to experiments the direct and indirect 
quasiparticle band gaps are overestimated when our EPI values are used.
Replacing our EPI contribution with the estimate by Gorelov {\em et. al.}
($\approx$-2~eV) yields underestimated band gaps compared to experiment.
From the relatively large difference between the EPI contributions computed in this work and Ref.~\cite{gorelov2020}, we
conclude that this contribution is the remaining leading order error in our study.
However, the experimental metallization pressure of about 450~GPa lies within our theoretical uncertainties.

\begin{figure}[ht]
\begin{center}
\includegraphics[width=0.99\linewidth]{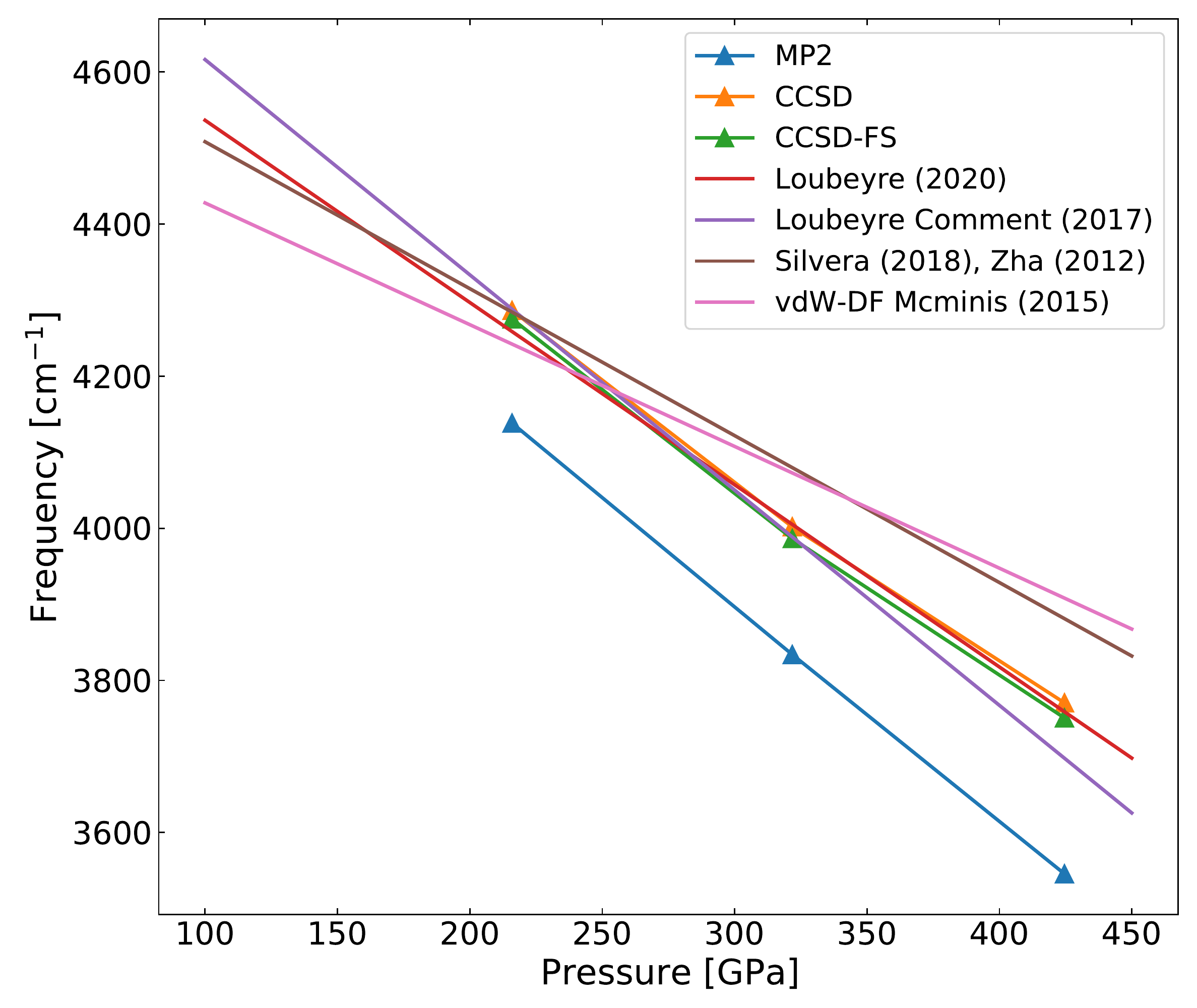}
\end{center}
\caption{\label{fig:freqvspressure}
Experimentally measured and theoretically calculated
H$_2$ vibron peak frequencies as a function of pressures.
The approximate vdW-DF (taken from the supplemental material of Ref.~\cite{mcminis2015}),
MP2 and CCSD harmonic frequencies are shown by
the pink, blue and green solid lines, respectively.
The brown solid line shows the experimentally measured relation between
the H$_2$ vibron frequencies and pressures from Silvera (2018)~\cite{silvera2018} and Zha
(2012)~\cite{zha2012}. Another two experimental data lines are from Loubeyre
(2020)~\cite{loubeyre2020} (red) and Loubeyre (2017)~\cite{loubeyre2017} (purple).
}
\end{figure}

For a deeper understanding of the comparison with experiments, we also assess the reliability 
of the experimental pressure calibration.
This is done by analyzing the dependence of the H$_2$ vibron peak frequency as a function of the
pressure.
As pointed out in Refs.~\cite{silvera2018,silvera2019,loubeyre2020} and depicted in Fig.~\ref{fig:freqvspressure},
the currently available experimental estimates for the H$_2$ vibron peak frequency vary
significantly at high pressures, questioning the reliability of experimentally determined pressures.
Possible reasons for the experimental uncertainties are summarised in
Ref.~\cite{silvera2019}.
However, theoretical estimates of the vibron peak frequency with respect to pressure
also vary significantly with respect to the employed XC parametrization on
the level of DFT~\cite{mcminis2015,loubeyre2020}.
We have estimated the vibrational frequency for the MP2 structures by computing the MP2 and CCSD
energies as a function of H$_2$ bond lengths around the equilibrium. Molecular orientations, locations of the centers of mass and
volumes are fixed while changing the bond lengths in accordance with Ref.~\cite{mcminis2015}.
The change of the harmonic frequency with respect to the pressure can be
used as a reliable calibration for pressures depicted in
Fig.~\ref{fig:freqvspressure}.
We find that both the MP2 and CCSD frequencies have a similar slope as the H$_2$
vibron frequency peak measured by Loubeyre {\em et. al.} in Ref.~\cite{loubeyre2020,loubeyre2017}.
From this we conclude that the experimental band gaps depicted Fig.~\ref{fig:gapvspressure} correspond to pressures that
are in good agreement with our most accurate estimates. In passing we note that despite the good agreement of vdW-DF structures
with our MP2 structures, vdW-DF vibrational frequencies are in better agreement with experimental results of Ref.~\cite{silvera2018,zha2012}.
However, we argue that this agreement is most likely fortuitous, because both MP2 and the more accurate CCSD vibrational
frequencies exhibit a very similar and steeper slope with respect to pressure.
From the above findings, we conclude that the vibrational frequencies of high pressure hydrogen phases are very sensitive
to the structural parameters and the corresponding electronic structure method.
This has potential implications for estimates of the zero point motion energy contribution to the lattice enthalpies
of accurate {\em ab initio} calculations of transition pressures~\cite{drummond2015}.
Having established the good agreement between our pressure estimates and those reported in Ref.~\cite{loubeyre2020},
we can also comment on the observed evidence of a phase transition at 425~GPa.
As predicted by both DMC and CCSD calculations~\cite{drummond2015,mcminis2015,liao2019} at low temperature,
phase III (C2/c-24) transforms into Cmca-12 at this pressure. However, these calculations have been performed
using DFT optimized structures. We have investigated the lowering of static lattice enthalpies resulting from
MP2 lattice relaxations for both structures at a selected volume corresponding to the
DFT pressure of 450~GPa, finding that changes to the previously calculated
transition pressures are negligible. This is surprising given the relatively large changes to the H$_2$ bond lengths.

{\bf Conclusions.}
Our work demonstrates the strengths and weaknesses of widely-used approximate
DFT methods for simulating high pressure phases of hydrogen by comparing to more accurate results obtained
using many-electron methods including coupled cluster theory.
Although approximate density functional theory is a computationally efficient tool for performing
random structure searches~\cite{pickard2007}, further structural optimization is required
to achieve good agreement of band gaps and vibrational frequencies with experimental findings in solid hydrogen.
Here, we demonstrate that periodic many-electron perturbation theory calculations using plane wave basis sets have become increasingly efficient
in previous years~\cite{gruber2018,liao2019}, making such optimisations feasible for systems with an increasing number of atoms.
Our findings show that compared to MP2 theory, DFT-PBE
structures exhibit too large hydrogen bond lengths causing too small band gaps.
Although vdW-DF calculations predict structures that are closer to MP2 theory,
vibrational frequencies that agree with experiment for a wide range
of pressures can only be obtained on the level of CCSD.
Furthermore we have demonstrated that the remaining leading order error of {\em ab initio} band gaps in solid
hydrogen crystals is likely to originate from approximations used to estimate the EPI contributions.
Nevertheless, it is worth pointing out that $T$-dependent 
fundamental band gap renormalization based on DFT-PBE structure is in better
agreement with the experimental data.
Combining accurate benchmark results with hybrid or non-local XC functionals using adjustable parameters could
be useful for materials modeling in this case.
Alternatively, machine-learning from MP2 forces or even more accurate {\em ab initio} data could be
used to produce accurate potential energy surfaces and corresponding vibrational entropy contributions.
Future work will focus on a seamless integration of electron-phonon interaction on the level of
many-electron theories to further improve the accuracy of such {\it ab initio} simulations.

{\bf Acknowledgments} K.L. thanks Daniel Kats for useful discussions
in the early stage of the project and Max Planck Institute for
Solid State Research for the PhD funding.
T.S. wants to thank Xiao-Wei Zhang and Qi-Jun Ye for fruitful discussions.
Supports and fundings from the European
Research Council (ERC) under the European Unions
Horizon 2020 research and innovation program (Grant Agreement No 715594), 
National Basic Research Programs of China under Grant 
Nos.~2016YFA0300900 and 2017YFA0205003, and
the National Science Foundation of China under Grant 
Nos.~11774003, 11934003, 11634001, and 21673005
are gratefully acknowledged.
The computational results presented have been achieved using
the Vienna Scientific Cluster (VSC), the cluster of the
Electronic Structure Theory Department at Max Planck Institute for
Solid State Research and the High-performance Computing Platform 
of Peking University, China.

{\bf Competing Interests} The authors declare that they have no
competing financial interests.

{\bf Correspondence} Correspondence and requests for materials
should be addressed to A.G.~(email:~andreas.grueneis@tuwien.ac.at).

\bibliography{article_arxiv}

\begin{thebibliography}{61}%
\makeatletter
\providecommand \@ifxundefined [1]{%
 \@ifx{#1\undefined}
}%
\providecommand \@ifnum [1]{%
 \ifnum #1\expandafter \@firstoftwo
 \else \expandafter \@secondoftwo
 \fi
}%
\providecommand \@ifx [1]{%
 \ifx #1\expandafter \@firstoftwo
 \else \expandafter \@secondoftwo
 \fi
}%
\providecommand \natexlab [1]{#1}%
\providecommand \enquote  [1]{``#1''}%
\providecommand \bibnamefont  [1]{#1}%
\providecommand \bibfnamefont [1]{#1}%
\providecommand \citenamefont [1]{#1}%
\providecommand \href@noop [0]{\@secondoftwo}%
\providecommand \href [0]{\begingroup \@sanitize@url \@href}%
\providecommand \@href[1]{\@@startlink{#1}\@@href}%
\providecommand \@@href[1]{\endgroup#1\@@endlink}%
\providecommand \@sanitize@url [0]{\catcode `\\12\catcode `\$12\catcode
  `\&12\catcode `\#12\catcode `\^12\catcode `\_12\catcode `\%12\relax}%
\providecommand \@@startlink[1]{}%
\providecommand \@@endlink[0]{}%
\providecommand \url  [0]{\begingroup\@sanitize@url \@url }%
\providecommand \@url [1]{\endgroup\@href {#1}{\urlprefix }}%
\providecommand \urlprefix  [0]{URL }%
\providecommand \Eprint [0]{\href }%
\providecommand \doibase [0]{http://dx.doi.org/}%
\providecommand \selectlanguage [0]{\@gobble}%
\providecommand \bibinfo  [0]{\@secondoftwo}%
\providecommand \bibfield  [0]{\@secondoftwo}%
\providecommand \translation [1]{[#1]}%
\providecommand \BibitemOpen [0]{}%
\providecommand \bibitemStop [0]{}%
\providecommand \bibitemNoStop [0]{.\EOS\space}%
\providecommand \EOS [0]{\spacefactor3000\relax}%
\providecommand \BibitemShut  [1]{\csname bibitem#1\endcsname}%
\let\auto@bib@innerbib\@empty
\bibitem [{\citenamefont {Wigner}\ and\ \citenamefont
  {Huntington}(1935)}]{wigner1935}%
  \BibitemOpen
  \bibfield  {author} {\bibinfo {author} {\bibfnamefont {E.}~\bibnamefont
  {Wigner}}\ and\ \bibinfo {author} {\bibfnamefont {H.~B.}\ \bibnamefont
  {Huntington}},\ }\href {http://aip.scitation.org/doi/10.1063/1.1749590}
  {\bibfield  {journal} {\bibinfo  {journal} {The Journal of Chemical Physics}\
  }\textbf {\bibinfo {volume} {3}},\ \bibinfo {pages} {764} (\bibinfo {year}
  {1935})},\ \bibinfo {note} {publisher: American Institute of Physics ISBN:
  0435160435}\BibitemShut {NoStop}%
\bibitem [{\citenamefont {Hemley}\ and\ \citenamefont
  {Mao}(1989)}]{hemley1989}%
  \BibitemOpen
  \bibfield  {author} {\bibinfo {author} {\bibfnamefont {R.~J.}\ \bibnamefont
  {Hemley}}\ and\ \bibinfo {author} {\bibfnamefont {H.-k.}\ \bibnamefont
  {Mao}},\ }\href {http://science.sciencemag.org/content/244/4911/1462}
  {\bibfield  {journal} {\bibinfo  {journal} {Science}\ }\textbf {\bibinfo
  {volume} {244}},\ \bibinfo {pages} {1462} (\bibinfo {year}
  {1989})}\BibitemShut {NoStop}%
\bibitem [{\citenamefont {Narayana}\ \emph {et~al.}(1998)\citenamefont
  {Narayana}, \citenamefont {Luo}, \citenamefont {Orloff},\ and\ \citenamefont
  {Ruoff}}]{narayana1998}%
  \BibitemOpen
  \bibfield  {author} {\bibinfo {author} {\bibfnamefont {C.}~\bibnamefont
  {Narayana}}, \bibinfo {author} {\bibfnamefont {H.}~\bibnamefont {Luo}},
  \bibinfo {author} {\bibfnamefont {J.}~\bibnamefont {Orloff}}, \ and\ \bibinfo
  {author} {\bibfnamefont {A.~L.}\ \bibnamefont {Ruoff}},\ }\href
  {http://www.nature.com/articles/29949} {\bibfield  {journal} {\bibinfo
  {journal} {Nature}\ }\textbf {\bibinfo {volume} {393}},\ \bibinfo {pages}
  {46} (\bibinfo {year} {1998})},\ \bibinfo {note} {arXiv: physics/9810036
  Publisher: Nature Publishing Group ISBN: 0028-0836}\BibitemShut {NoStop}%
\bibitem [{\citenamefont {Loubeyre}\ \emph {et~al.}(2002)\citenamefont
  {Loubeyre}, \citenamefont {Occelli},\ and\ \citenamefont
  {LeToullec}}]{loubeyre2002}%
  \BibitemOpen
  \bibfield  {author} {\bibinfo {author} {\bibfnamefont {P.}~\bibnamefont
  {Loubeyre}}, \bibinfo {author} {\bibfnamefont {F.}~\bibnamefont {Occelli}}, \
  and\ \bibinfo {author} {\bibfnamefont {R.}~\bibnamefont {LeToullec}},\ }\href
  {http://www.nature.com/articles/416613a} {\bibfield  {journal} {\bibinfo
  {journal} {Nature}\ }\textbf {\bibinfo {volume} {416}},\ \bibinfo {pages}
  {613} (\bibinfo {year} {2002})},\ \bibinfo {note} {publisher: Nature
  Publishing Group ISBN: 0028-0836}\BibitemShut {NoStop}%
\bibitem [{\citenamefont {Dias}\ and\ \citenamefont
  {Silvera}(2017{\natexlab{a}})}]{dias2017}%
  \BibitemOpen
  \bibfield  {author} {\bibinfo {author} {\bibfnamefont {R.~P.}\ \bibnamefont
  {Dias}}\ and\ \bibinfo {author} {\bibfnamefont {I.~F.}\ \bibnamefont
  {Silvera}},\ }\href {http://www.ncbi.nlm.nih.gov/pubmed/28818917} {\bibfield
  {journal} {\bibinfo  {journal} {Science}\ }\textbf {\bibinfo {volume}
  {357}},\ \bibinfo {pages} {eaao5843} (\bibinfo {year}
  {2017}{\natexlab{a}})},\ \bibinfo {note} {publisher: American Association for
  the Advancement of Science}\BibitemShut {NoStop}%
\bibitem [{\citenamefont {Dias}\ and\ \citenamefont
  {Silvera}(2017{\natexlab{b}})}]{dias2017a}%
  \BibitemOpen
  \bibfield  {author} {\bibinfo {author} {\bibfnamefont {R.~P.}\ \bibnamefont
  {Dias}}\ and\ \bibinfo {author} {\bibfnamefont {I.~F.}\ \bibnamefont
  {Silvera}},\ }\href {http://www.ncbi.nlm.nih.gov/pubmed/28126728} {\bibfield
  {journal} {\bibinfo  {journal} {Science}\ }\textbf {\bibinfo {volume}
  {355}},\ \bibinfo {pages} {715} (\bibinfo {year} {2017}{\natexlab{b}})},\
  \bibinfo {note} {arXiv: 1703.03064 Publisher: American Association for the
  Advancement of Science ISBN: 1095-9203 (Electronic) 0036-8075
  (Linking)}\BibitemShut {NoStop}%
\bibitem [{\citenamefont {Loubeyre}\ \emph {et~al.}(2020)\citenamefont
  {Loubeyre}, \citenamefont {Occelli},\ and\ \citenamefont
  {Dumas}}]{loubeyre2020}%
  \BibitemOpen
  \bibfield  {author} {\bibinfo {author} {\bibfnamefont {P.}~\bibnamefont
  {Loubeyre}}, \bibinfo {author} {\bibfnamefont {F.}~\bibnamefont {Occelli}}, \
  and\ \bibinfo {author} {\bibfnamefont {P.}~\bibnamefont {Dumas}},\ }\href
  {\doibase 10.1038/s41586-019-1927-3} {\bibfield  {journal} {\bibinfo
  {journal} {Nature}\ }\textbf {\bibinfo {volume} {577}},\ \bibinfo {pages}
  {631} (\bibinfo {year} {2020})}\BibitemShut {NoStop}%
\bibitem [{\citenamefont {Geng}(2017)}]{geng2017}%
  \BibitemOpen
  \bibfield  {author} {\bibinfo {author} {\bibfnamefont {H.~Y.}\ \bibnamefont
  {Geng}},\ }\href
  {https://aip.scitation.org/doi/abs/10.1016/j.mre.2017.10.001} {\bibfield
  {journal} {\bibinfo  {journal} {Matter and Radiation at Extremes}\ }\textbf
  {\bibinfo {volume} {2}},\ \bibinfo {pages} {275} (\bibinfo {year}
  {2017})}\BibitemShut {NoStop}%
\bibitem [{\citenamefont {Silvera}\ and\ \citenamefont
  {Dias}(2019)}]{silvera2019}%
  \BibitemOpen
  \bibfield  {author} {\bibinfo {author} {\bibfnamefont {I.~F.}\ \bibnamefont
  {Silvera}}\ and\ \bibinfo {author} {\bibfnamefont {R.}~\bibnamefont {Dias}},\
  }\href {http://arxiv.org/abs/1907.03198} {\bibfield  {journal} {\bibinfo
  {journal} {arXiv:1907.03198 [cond-mat]}\ } (\bibinfo {year} {2019})},\
  \bibinfo {note} {arXiv: 1907.03198}\BibitemShut {NoStop}%
\bibitem [{\citenamefont {Hemley}\ and\ \citenamefont
  {Mao}(1988)}]{hemley1988}%
  \BibitemOpen
  \bibfield  {author} {\bibinfo {author} {\bibfnamefont {R.~J.}\ \bibnamefont
  {Hemley}}\ and\ \bibinfo {author} {\bibfnamefont {H.-k.}\ \bibnamefont
  {Mao}},\ }\href@noop {} {\emph {\bibinfo {title} {Phase {Transition} in
  {Solid} {Molecular} {Hydrogen} at {Ultrahigh} {Pressures}}}},\ \bibinfo
  {type} {Tech. Rep.}\ (\bibinfo {year} {1988})\ \bibinfo {note} {volume:
  61}\BibitemShut {NoStop}%
\bibitem [{\citenamefont {Lorenzana}\ \emph {et~al.}(1989)\citenamefont
  {Lorenzana}, \citenamefont {Silvera},\ and\ \citenamefont
  {Goettel}}]{lorenzana1989}%
  \BibitemOpen
  \bibfield  {author} {\bibinfo {author} {\bibfnamefont {H.~E.}\ \bibnamefont
  {Lorenzana}}, \bibinfo {author} {\bibfnamefont {I.~F.}\ \bibnamefont
  {Silvera}}, \ and\ \bibinfo {author} {\bibfnamefont {K.~A.}\ \bibnamefont
  {Goettel}},\ }\href {https://link.aps.org/doi/10.1103/PhysRevLett.63.2080}
  {\bibfield  {journal} {\bibinfo  {journal} {Physical Review Letters}\
  }\textbf {\bibinfo {volume} {63}},\ \bibinfo {pages} {2080} (\bibinfo {year}
  {1989})}\BibitemShut {NoStop}%
\bibitem [{\citenamefont {Mao}\ and\ \citenamefont {Hemley}(1994)}]{mao1994}%
  \BibitemOpen
  \bibfield  {author} {\bibinfo {author} {\bibfnamefont {H.~K.}\ \bibnamefont
  {Mao}}\ and\ \bibinfo {author} {\bibfnamefont {R.~J.}\ \bibnamefont
  {Hemley}},\ }\href@noop {} {\bibfield  {journal} {\bibinfo  {journal}
  {Reviews of Modern Physics}\ }\textbf {\bibinfo {volume} {66}},\ \bibinfo
  {pages} {671} (\bibinfo {year} {1994})},\ \bibinfo {note} {iSBN:
  0034-6861}\BibitemShut {NoStop}%
\bibitem [{\citenamefont {Natoli}\ \emph {et~al.}(1995)\citenamefont {Natoli},
  \citenamefont {Martin},\ and\ \citenamefont {Ceperley}}]{natoli1995}%
  \BibitemOpen
  \bibfield  {author} {\bibinfo {author} {\bibfnamefont {V.}~\bibnamefont
  {Natoli}}, \bibinfo {author} {\bibfnamefont {R.~M.}\ \bibnamefont {Martin}},
  \ and\ \bibinfo {author} {\bibfnamefont {D.}~\bibnamefont {Ceperley}},\
  }\href {https://link.aps.org/doi/10.1103/PhysRevLett.74.1601} {\bibfield
  {journal} {\bibinfo  {journal} {Physical Review Letters}\ }\textbf {\bibinfo
  {volume} {74}},\ \bibinfo {pages} {1601} (\bibinfo {year}
  {1995})}\BibitemShut {NoStop}%
\bibitem [{\citenamefont {Goncharenko}\ and\ \citenamefont
  {Loubeyre}(2005)}]{goncharenko2005}%
  \BibitemOpen
  \bibfield  {author} {\bibinfo {author} {\bibfnamefont {I.}~\bibnamefont
  {Goncharenko}}\ and\ \bibinfo {author} {\bibfnamefont {P.}~\bibnamefont
  {Loubeyre}},\ }\href {https://www.nature.com/articles/nature03699} {\bibfield
   {journal} {\bibinfo  {journal} {Nature}\ }\textbf {\bibinfo {volume}
  {435}},\ \bibinfo {pages} {1206} (\bibinfo {year} {2005})}\BibitemShut
  {NoStop}%
\bibitem [{\citenamefont {Pickard}\ and\ \citenamefont
  {Needs}(2007)}]{pickard2007}%
  \BibitemOpen
  \bibfield  {author} {\bibinfo {author} {\bibfnamefont {C.~J.}\ \bibnamefont
  {Pickard}}\ and\ \bibinfo {author} {\bibfnamefont {R.~J.}\ \bibnamefont
  {Needs}},\ }\href {http://www.nature.com/articles/nphys625} {\bibfield
  {journal} {\bibinfo  {journal} {Nature Physics}\ }\textbf {\bibinfo {volume}
  {3}},\ \bibinfo {pages} {473} (\bibinfo {year} {2007})},\ \bibinfo {note}
  {arXiv: 1609.07486 Publisher: Nature Publishing Group ISBN: 1745-2473
  1745-2481}\BibitemShut {NoStop}%
\bibitem [{\citenamefont {Howie}\ \emph {et~al.}(2012)\citenamefont {Howie},
  \citenamefont {Guillaume}, \citenamefont {Scheler}, \citenamefont
  {Goncharov},\ and\ \citenamefont {Gregoryanz}}]{howie2012}%
  \BibitemOpen
  \bibfield  {author} {\bibinfo {author} {\bibfnamefont {R.~T.}\ \bibnamefont
  {Howie}}, \bibinfo {author} {\bibfnamefont {C.~L.}\ \bibnamefont
  {Guillaume}}, \bibinfo {author} {\bibfnamefont {T.}~\bibnamefont {Scheler}},
  \bibinfo {author} {\bibfnamefont {A.~F.}\ \bibnamefont {Goncharov}}, \ and\
  \bibinfo {author} {\bibfnamefont {E.}~\bibnamefont {Gregoryanz}},\ }\href
  {https://link.aps.org/doi/10.1103/PhysRevLett.108.125501} {\bibfield
  {journal} {\bibinfo  {journal} {Physical Review Letters}\ }\textbf {\bibinfo
  {volume} {108}},\ \bibinfo {pages} {125501} (\bibinfo {year}
  {2012})}\BibitemShut {NoStop}%
\bibitem [{\citenamefont {Weir}\ \emph {et~al.}(1996)\citenamefont {Weir},
  \citenamefont {Mitchell},\ and\ \citenamefont {Nellis}}]{weir1996}%
  \BibitemOpen
  \bibfield  {author} {\bibinfo {author} {\bibfnamefont {S.~T.}\ \bibnamefont
  {Weir}}, \bibinfo {author} {\bibfnamefont {A.~C.}\ \bibnamefont {Mitchell}},
  \ and\ \bibinfo {author} {\bibfnamefont {W.~J.}\ \bibnamefont {Nellis}},\
  }\href {https://link.aps.org/doi/10.1103/PhysRevLett.76.1860} {\bibfield
  {journal} {\bibinfo  {journal} {Physical Review Letters}\ }\textbf {\bibinfo
  {volume} {76}},\ \bibinfo {pages} {1860} (\bibinfo {year} {1996})},\ \bibinfo
  {note} {publisher: American Physical Society}\BibitemShut {NoStop}%
\bibitem [{\citenamefont {Johnson}\ and\ \citenamefont
  {Ashcroft}(2000)}]{johnson2000}%
  \BibitemOpen
  \bibfield  {author} {\bibinfo {author} {\bibfnamefont {K.~A.}\ \bibnamefont
  {Johnson}}\ and\ \bibinfo {author} {\bibfnamefont {N.~W.}\ \bibnamefont
  {Ashcroft}},\ }\href@noop {} {\bibfield  {journal} {\bibinfo  {journal}
  {Nature}\ }\textbf {\bibinfo {volume} {403}},\ \bibinfo {pages} {632}
  (\bibinfo {year} {2000})},\ \bibinfo {note} {iSBN: 0028-0836}\BibitemShut
  {NoStop}%
\bibitem [{\citenamefont {Eremets}\ and\ \citenamefont
  {Troyan}(2011)}]{eremets2011}%
  \BibitemOpen
  \bibfield  {author} {\bibinfo {author} {\bibfnamefont {M.~I.}\ \bibnamefont
  {Eremets}}\ and\ \bibinfo {author} {\bibfnamefont {I.~A.}\ \bibnamefont
  {Troyan}},\ }\href {http://www.nature.com/articles/nmat3175} {\bibfield
  {journal} {\bibinfo  {journal} {Nature Materials}\ }\textbf {\bibinfo
  {volume} {10}},\ \bibinfo {pages} {927} (\bibinfo {year} {2011})},\ \bibinfo
  {note} {publisher: Nature Publishing Group ISBN: 1476-1122}\BibitemShut
  {NoStop}%
\bibitem [{\citenamefont {Chen}\ \emph {et~al.}(2013)\citenamefont {Chen},
  \citenamefont {Li}, \citenamefont {Zhang}, \citenamefont {Probert},
  \citenamefont {Pickard}, \citenamefont {Needs}, \citenamefont {Michaelides},\
  and\ \citenamefont {Wang}}]{chen2013}%
  \BibitemOpen
  \bibfield  {author} {\bibinfo {author} {\bibfnamefont {J.}~\bibnamefont
  {Chen}}, \bibinfo {author} {\bibfnamefont {X.-Z.}\ \bibnamefont {Li}},
  \bibinfo {author} {\bibfnamefont {Q.}~\bibnamefont {Zhang}}, \bibinfo
  {author} {\bibfnamefont {M.~I.~J.}\ \bibnamefont {Probert}}, \bibinfo
  {author} {\bibfnamefont {C.~J.}\ \bibnamefont {Pickard}}, \bibinfo {author}
  {\bibfnamefont {R.~J.}\ \bibnamefont {Needs}}, \bibinfo {author}
  {\bibfnamefont {A.}~\bibnamefont {Michaelides}}, \ and\ \bibinfo {author}
  {\bibfnamefont {E.}~\bibnamefont {Wang}},\ }\href
  {https://www.nature.com/articles/ncomms3064} {\bibfield  {journal} {\bibinfo
  {journal} {Nature Communications}\ }\textbf {\bibinfo {volume} {4}},\
  \bibinfo {pages} {2064} (\bibinfo {year} {2013})}\BibitemShut {NoStop}%
\bibitem [{\citenamefont {Knudson}\ \emph {et~al.}(2015)\citenamefont
  {Knudson}, \citenamefont {Desjarlais}, \citenamefont {Becker}, \citenamefont
  {Lemke}, \citenamefont {Cochrane}, \citenamefont {Savage}, \citenamefont
  {Bliss}, \citenamefont {Mattsson},\ and\ \citenamefont
  {Redmer}}]{knudson2015}%
  \BibitemOpen
  \bibfield  {author} {\bibinfo {author} {\bibfnamefont {M.~D.}\ \bibnamefont
  {Knudson}}, \bibinfo {author} {\bibfnamefont {M.~P.}\ \bibnamefont
  {Desjarlais}}, \bibinfo {author} {\bibfnamefont {A.}~\bibnamefont {Becker}},
  \bibinfo {author} {\bibfnamefont {R.~W.}\ \bibnamefont {Lemke}}, \bibinfo
  {author} {\bibfnamefont {K.~R.}\ \bibnamefont {Cochrane}}, \bibinfo {author}
  {\bibfnamefont {M.~E.}\ \bibnamefont {Savage}}, \bibinfo {author}
  {\bibfnamefont {D.~E.}\ \bibnamefont {Bliss}}, \bibinfo {author}
  {\bibfnamefont {T.~R.}\ \bibnamefont {Mattsson}}, \ and\ \bibinfo {author}
  {\bibfnamefont {R.}~\bibnamefont {Redmer}},\ }\href
  {http://science.sciencemag.org/content/348/6242/1455} {\bibfield  {journal}
  {\bibinfo  {journal} {Science}\ }\textbf {\bibinfo {volume} {348}},\ \bibinfo
  {pages} {1455} (\bibinfo {year} {2015})}\BibitemShut {NoStop}%
\bibitem [{\citenamefont {Zaghoo}\ \emph {et~al.}(2016)\citenamefont {Zaghoo},
  \citenamefont {Salamat},\ and\ \citenamefont {Silvera}}]{zaghoo2016}%
  \BibitemOpen
  \bibfield  {author} {\bibinfo {author} {\bibfnamefont {M.}~\bibnamefont
  {Zaghoo}}, \bibinfo {author} {\bibfnamefont {A.}~\bibnamefont {Salamat}}, \
  and\ \bibinfo {author} {\bibfnamefont {I.~F.}\ \bibnamefont {Silvera}},\
  }\href {https://link.aps.org/doi/10.1103/PhysRevB.93.155128} {\bibfield
  {journal} {\bibinfo  {journal} {Physical Review B}\ }\textbf {\bibinfo
  {volume} {93}},\ \bibinfo {pages} {155128} (\bibinfo {year}
  {2016})}\BibitemShut {NoStop}%
\bibitem [{\citenamefont {Li}\ \emph {et~al.}(2013)\citenamefont {Li},
  \citenamefont {Walker}, \citenamefont {Probert}, \citenamefont {Pickard},
  \citenamefont {Needs},\ and\ \citenamefont {Michaelides}}]{li2013}%
  \BibitemOpen
  \bibfield  {author} {\bibinfo {author} {\bibfnamefont {X.-Z.}\ \bibnamefont
  {Li}}, \bibinfo {author} {\bibfnamefont {B.}~\bibnamefont {Walker}}, \bibinfo
  {author} {\bibfnamefont {M.~I.~J.}\ \bibnamefont {Probert}}, \bibinfo
  {author} {\bibfnamefont {C.~J.}\ \bibnamefont {Pickard}}, \bibinfo {author}
  {\bibfnamefont {R.~J.}\ \bibnamefont {Needs}}, \ and\ \bibinfo {author}
  {\bibfnamefont {A.}~\bibnamefont {Michaelides}},\ }\href
  {http://www.ncbi.nlm.nih.gov/pubmed/23360786} {\bibfield  {journal} {\bibinfo
   {journal} {Journal of physics: Condensed Matter}\ }\textbf {\bibinfo
  {volume} {25}},\ \bibinfo {pages} {085402} (\bibinfo {year} {2013})},\
  \bibinfo {note} {arXiv: 1302.0062v1 ISBN: 0953-8984}\BibitemShut {NoStop}%
\bibitem [{\citenamefont {Drummond}\ \emph {et~al.}(2015)\citenamefont
  {Drummond}, \citenamefont {Monserrat}, \citenamefont {Lloyd-Williams},
  \citenamefont {Ríos}, \citenamefont {Pickard},\ and\ \citenamefont
  {Needs}}]{drummond2015}%
  \BibitemOpen
  \bibfield  {author} {\bibinfo {author} {\bibfnamefont {N.~D.}\ \bibnamefont
  {Drummond}}, \bibinfo {author} {\bibfnamefont {B.}~\bibnamefont {Monserrat}},
  \bibinfo {author} {\bibfnamefont {J.~H.}\ \bibnamefont {Lloyd-Williams}},
  \bibinfo {author} {\bibfnamefont {P.~L.}\ \bibnamefont {Ríos}}, \bibinfo
  {author} {\bibfnamefont {C.~J.}\ \bibnamefont {Pickard}}, \ and\ \bibinfo
  {author} {\bibfnamefont {R.~J.}\ \bibnamefont {Needs}},\ }\href
  {http://www.nature.com/doifinder/10.1038/ncomms8794} {\bibfield  {journal}
  {\bibinfo  {journal} {Nature Communications}\ }\textbf {\bibinfo {volume}
  {6}},\ \bibinfo {pages} {7794} (\bibinfo {year} {2015})}\BibitemShut
  {NoStop}%
\bibitem [{\citenamefont {Azadi}\ \emph {et~al.}(2013)\citenamefont {Azadi},
  \citenamefont {Foulkes},\ and\ \citenamefont {K\"uhne}}]{azadi2013}%
  \BibitemOpen
  \bibfield  {author} {\bibinfo {author} {\bibfnamefont {S.}~\bibnamefont
  {Azadi}}, \bibinfo {author} {\bibfnamefont {W.~M.~C.}\ \bibnamefont
  {Foulkes}}, \ and\ \bibinfo {author} {\bibfnamefont {T.~D.}\ \bibnamefont
  {K\"uhne}},\ }\href
  {http://stacks.iop.org/1367-2630/15/i=11/a=113005?key=crossref.bae4c993466b7ba0ae94ddcb62ff6b95}
  {\bibfield  {journal} {\bibinfo  {journal} {New Journal of Physics}\ }\textbf
  {\bibinfo {volume} {15}},\ \bibinfo {pages} {113005} (\bibinfo {year}
  {2013})},\ \bibinfo {note} {arXiv: 1307.1463}\BibitemShut {NoStop}%
\bibitem [{\citenamefont {McMinis}\ \emph {et~al.}(2015)\citenamefont
  {McMinis}, \citenamefont {Clay}, \citenamefont {Lee},\ and\ \citenamefont
  {Morales}}]{mcminis2015}%
  \BibitemOpen
  \bibfield  {author} {\bibinfo {author} {\bibfnamefont {J.}~\bibnamefont
  {McMinis}}, \bibinfo {author} {\bibfnamefont {R.~C.}\ \bibnamefont {Clay}},
  \bibinfo {author} {\bibfnamefont {D.}~\bibnamefont {Lee}}, \ and\ \bibinfo
  {author} {\bibfnamefont {M.~A.}\ \bibnamefont {Morales}},\ }\href@noop {}
  {\bibfield  {journal} {\bibinfo  {journal} {Physical Review Letters}\
  }\textbf {\bibinfo {volume} {114}} (\bibinfo {year} {2015})}\BibitemShut
  {NoStop}%
\bibitem [{\citenamefont {Morales}\ \emph
  {et~al.}(2013{\natexlab{a}})\citenamefont {Morales}, \citenamefont {McMahon},
  \citenamefont {Pierleoni},\ and\ \citenamefont {Ceperley}}]{morales2013}%
  \BibitemOpen
  \bibfield  {author} {\bibinfo {author} {\bibfnamefont {M.~A.}\ \bibnamefont
  {Morales}}, \bibinfo {author} {\bibfnamefont {J.~M.}\ \bibnamefont
  {McMahon}}, \bibinfo {author} {\bibfnamefont {C.}~\bibnamefont {Pierleoni}},
  \ and\ \bibinfo {author} {\bibfnamefont {D.~M.}\ \bibnamefont {Ceperley}},\
  }\href {https://link.aps.org/doi/10.1103/PhysRevB.87.184107} {\bibfield
  {journal} {\bibinfo  {journal} {Physical Review B}\ }\textbf {\bibinfo
  {volume} {87}},\ \bibinfo {pages} {184107} (\bibinfo {year}
  {2013}{\natexlab{a}})}\BibitemShut {NoStop}%
\bibitem [{\citenamefont {Morales}\ \emph
  {et~al.}(2013{\natexlab{b}})\citenamefont {Morales}, \citenamefont {McMahon},
  \citenamefont {Pierleoni},\ and\ \citenamefont {Ceperley}}]{morales2013a}%
  \BibitemOpen
  \bibfield  {author} {\bibinfo {author} {\bibfnamefont {M.~A.}\ \bibnamefont
  {Morales}}, \bibinfo {author} {\bibfnamefont {J.~M.}\ \bibnamefont
  {McMahon}}, \bibinfo {author} {\bibfnamefont {C.}~\bibnamefont {Pierleoni}},
  \ and\ \bibinfo {author} {\bibfnamefont {D.~M.}\ \bibnamefont {Ceperley}},\
  }\href {https://link.aps.org/doi/10.1103/PhysRevLett.110.065702} {\bibfield
  {journal} {\bibinfo  {journal} {Physical Review Letters}\ }\textbf {\bibinfo
  {volume} {110}},\ \bibinfo {pages} {065702} (\bibinfo {year}
  {2013}{\natexlab{b}})}\BibitemShut {NoStop}%
\bibitem [{\citenamefont {Chacham}\ and\ \citenamefont
  {Louie}(1991)}]{chacham1991}%
  \BibitemOpen
  \bibfield  {author} {\bibinfo {author} {\bibfnamefont {H.}~\bibnamefont
  {Chacham}}\ and\ \bibinfo {author} {\bibfnamefont {S.~G.}\ \bibnamefont
  {Louie}},\ }\href {https://link.aps.org/doi/10.1103/PhysRevLett.66.64}
  {\bibfield  {journal} {\bibinfo  {journal} {Physical Review Letters}\
  }\textbf {\bibinfo {volume} {66}},\ \bibinfo {pages} {64} (\bibinfo {year}
  {1991})},\ \bibinfo {note} {publisher: American Physical Society}\BibitemShut
  {NoStop}%
\bibitem [{\citenamefont {Kohanoff}\ \emph {et~al.}(1997)\citenamefont
  {Kohanoff}, \citenamefont {Scandolo}, \citenamefont {Chiarotti},\ and\
  \citenamefont {Tosatti}}]{kohanoff1997}%
  \BibitemOpen
  \bibfield  {author} {\bibinfo {author} {\bibfnamefont {J.}~\bibnamefont
  {Kohanoff}}, \bibinfo {author} {\bibfnamefont {S.}~\bibnamefont {Scandolo}},
  \bibinfo {author} {\bibfnamefont {G.~L.}\ \bibnamefont {Chiarotti}}, \ and\
  \bibinfo {author} {\bibfnamefont {E.}~\bibnamefont {Tosatti}},\ }\href
  {https://link.aps.org/doi/10.1103/PhysRevLett.78.2783} {\bibfield  {journal}
  {\bibinfo  {journal} {Physical Review Letters}\ }\textbf {\bibinfo {volume}
  {78}},\ \bibinfo {pages} {2783} (\bibinfo {year} {1997})}\BibitemShut
  {NoStop}%
\bibitem [{\citenamefont {Ceperley}\ and\ \citenamefont
  {Alder}(1987)}]{ceperley1987}%
  \BibitemOpen
  \bibfield  {author} {\bibinfo {author} {\bibfnamefont {D.~M.}\ \bibnamefont
  {Ceperley}}\ and\ \bibinfo {author} {\bibfnamefont {B.~J.}\ \bibnamefont
  {Alder}},\ }\href {https://link.aps.org/doi/10.1103/PhysRevB.36.2092}
  {\bibfield  {journal} {\bibinfo  {journal} {Physical Review B}\ }\textbf
  {\bibinfo {volume} {36}},\ \bibinfo {pages} {2092} (\bibinfo {year}
  {1987})},\ \bibinfo {note} {publisher: American Physical Society}\BibitemShut
  {NoStop}%
\bibitem [{\citenamefont {Clay}\ \emph {et~al.}(2014)\citenamefont {Clay},
  \citenamefont {McMinis}, \citenamefont {McMahon}, \citenamefont {Pierleoni},
  \citenamefont {Ceperley},\ and\ \citenamefont {Morales}}]{clay2014}%
  \BibitemOpen
  \bibfield  {author} {\bibinfo {author} {\bibfnamefont {R.~C.}\ \bibnamefont
  {Clay}}, \bibinfo {author} {\bibfnamefont {J.}~\bibnamefont {McMinis}},
  \bibinfo {author} {\bibfnamefont {J.~M.}\ \bibnamefont {McMahon}}, \bibinfo
  {author} {\bibfnamefont {C.}~\bibnamefont {Pierleoni}}, \bibinfo {author}
  {\bibfnamefont {D.~M.}\ \bibnamefont {Ceperley}}, \ and\ \bibinfo {author}
  {\bibfnamefont {M.~A.}\ \bibnamefont {Morales}},\ }\href@noop {} {\bibfield
  {journal} {\bibinfo  {journal} {Physical Review B - Condensed Matter and
  Materials Physics}\ }\textbf {\bibinfo {volume} {89}},\ \bibinfo {pages}
  {184106} (\bibinfo {year} {2014})},\ \bibinfo {note} {arXiv:
  1401.7365}\BibitemShut {NoStop}%
\bibitem [{\citenamefont {Azadi}\ \emph {et~al.}(2014)\citenamefont {Azadi},
  \citenamefont {Monserrat}, \citenamefont {Foulkes},\ and\ \citenamefont
  {Needs}}]{azadi2014}%
  \BibitemOpen
  \bibfield  {author} {\bibinfo {author} {\bibfnamefont {S.}~\bibnamefont
  {Azadi}}, \bibinfo {author} {\bibfnamefont {B.}~\bibnamefont {Monserrat}},
  \bibinfo {author} {\bibfnamefont {W.~M.~C.}\ \bibnamefont {Foulkes}}, \ and\
  \bibinfo {author} {\bibfnamefont {R.~J.}\ \bibnamefont {Needs}},\ }\href@noop
  {} {\bibfield  {journal} {\bibinfo  {journal} {Physical Review Letters}\
  }\textbf {\bibinfo {volume} {112}} (\bibinfo {year} {2014})},\ \bibinfo
  {note} {arXiv: 1403.3681 ISBN: 0031-9007}\BibitemShut {NoStop}%
\bibitem [{\citenamefont {Azadi}\ \emph {et~al.}(2017)\citenamefont {Azadi},
  \citenamefont {Drummond},\ and\ \citenamefont {Foulkes}}]{azadi2017}%
  \BibitemOpen
  \bibfield  {author} {\bibinfo {author} {\bibfnamefont {S.}~\bibnamefont
  {Azadi}}, \bibinfo {author} {\bibfnamefont {N.~D.}\ \bibnamefont {Drummond}},
  \ and\ \bibinfo {author} {\bibfnamefont {W.~M.~C.}\ \bibnamefont {Foulkes}},\
  }\href@noop {} {\bibfield  {journal} {\bibinfo  {journal} {Physical Review
  B}\ }\textbf {\bibinfo {volume} {95}},\ \bibinfo {pages} {35142} (\bibinfo
  {year} {2017})},\ \bibinfo {note} {arXiv: 1608.00754}\BibitemShut {NoStop}%
\bibitem [{\citenamefont {Zha}\ \emph {et~al.}(2012)\citenamefont {Zha},
  \citenamefont {Liu},\ and\ \citenamefont {Hemley}}]{zha2012}%
  \BibitemOpen
  \bibfield  {author} {\bibinfo {author} {\bibfnamefont {C.-S.}\ \bibnamefont
  {Zha}}, \bibinfo {author} {\bibfnamefont {Z.}~\bibnamefont {Liu}}, \ and\
  \bibinfo {author} {\bibfnamefont {R.~J.}\ \bibnamefont {Hemley}},\ }\href
  {https://link.aps.org/doi/10.1103/PhysRevLett.108.146402} {\bibfield
  {journal} {\bibinfo  {journal} {Physical Review Letters}\ }\textbf {\bibinfo
  {volume} {108}},\ \bibinfo {pages} {146402} (\bibinfo {year}
  {2012})}\BibitemShut {NoStop}%
\bibitem [{\citenamefont {Hanfland}\ \emph {et~al.}(1993)\citenamefont
  {Hanfland}, \citenamefont {Hemley},\ and\ \citenamefont
  {Mao}}]{hanfland1993}%
  \BibitemOpen
  \bibfield  {author} {\bibinfo {author} {\bibfnamefont {M.}~\bibnamefont
  {Hanfland}}, \bibinfo {author} {\bibfnamefont {R.~J.}\ \bibnamefont
  {Hemley}}, \ and\ \bibinfo {author} {\bibfnamefont {H.-k.}\ \bibnamefont
  {Mao}},\ }\href {https://link.aps.org/doi/10.1103/PhysRevLett.70.3760}
  {\bibfield  {journal} {\bibinfo  {journal} {Physical Review Letters}\
  }\textbf {\bibinfo {volume} {70}},\ \bibinfo {pages} {3760} (\bibinfo {year}
  {1993})}\BibitemShut {NoStop}%
\bibitem [{\citenamefont {Monserrat}\ \emph {et~al.}(2016)\citenamefont
  {Monserrat}, \citenamefont {Needs}, \citenamefont {Gregoryanz},\ and\
  \citenamefont {Pickard}}]{monserrat2016}%
  \BibitemOpen
  \bibfield  {author} {\bibinfo {author} {\bibfnamefont {B.}~\bibnamefont
  {Monserrat}}, \bibinfo {author} {\bibfnamefont {R.~J.}\ \bibnamefont
  {Needs}}, \bibinfo {author} {\bibfnamefont {E.}~\bibnamefont {Gregoryanz}}, \
  and\ \bibinfo {author} {\bibfnamefont {C.~J.}\ \bibnamefont {Pickard}},\
  }\href {https://link.aps.org/doi/10.1103/PhysRevB.94.134101} {\bibfield
  {journal} {\bibinfo  {journal} {Physical Review B}\ }\textbf {\bibinfo
  {volume} {94}},\ \bibinfo {pages} {134101} (\bibinfo {year}
  {2016})}\BibitemShut {NoStop}%
\bibitem [{\citenamefont {Zhang}\ \emph {et~al.}(2018)\citenamefont {Zhang},
  \citenamefont {Wang},\ and\ \citenamefont {Li}}]{zhang2018}%
  \BibitemOpen
  \bibfield  {author} {\bibinfo {author} {\bibfnamefont {X.-W.}\ \bibnamefont
  {Zhang}}, \bibinfo {author} {\bibfnamefont {E.-G.}\ \bibnamefont {Wang}}, \
  and\ \bibinfo {author} {\bibfnamefont {X.-Z.}\ \bibnamefont {Li}},\ }\href
  {https://link.aps.org/doi/10.1103/PhysRevB.98.134110} {\bibfield  {journal}
  {\bibinfo  {journal} {Physical Review B}\ }\textbf {\bibinfo {volume} {98}},\
  \bibinfo {pages} {134110} (\bibinfo {year} {2018})}\BibitemShut {NoStop}%
\bibitem [{\citenamefont {Azadi}\ and\ \citenamefont
  {Ackland}(2017)}]{azadi2017c}%
  \BibitemOpen
  \bibfield  {author} {\bibinfo {author} {\bibfnamefont {S.}~\bibnamefont
  {Azadi}}\ and\ \bibinfo {author} {\bibfnamefont {G.~J.}\ \bibnamefont
  {Ackland}},\ }\href
  {https://pubs.rsc.org/en/content/articlelanding/2017/cp/c7cp03729e}
  {\bibfield  {journal} {\bibinfo  {journal} {Physical Chemistry Chemical
  Physics}\ }\textbf {\bibinfo {volume} {19}},\ \bibinfo {pages} {21829}
  (\bibinfo {year} {2017})}\BibitemShut {NoStop}%
\bibitem [{\citenamefont {Gorelov}\ \emph {et~al.}(2020)\citenamefont
  {Gorelov}, \citenamefont {Holzmann}, \citenamefont {Ceperley},\ and\
  \citenamefont {Pierleoni}}]{gorelov2020}%
  \BibitemOpen
  \bibfield  {author} {\bibinfo {author} {\bibfnamefont {V.}~\bibnamefont
  {Gorelov}}, \bibinfo {author} {\bibfnamefont {M.}~\bibnamefont {Holzmann}},
  \bibinfo {author} {\bibfnamefont {D.~M.}\ \bibnamefont {Ceperley}}, \ and\
  \bibinfo {author} {\bibfnamefont {C.}~\bibnamefont {Pierleoni}},\ }\href
  {\doibase 10.1103/physrevlett.124.116401} {\bibfield  {journal} {\bibinfo
  {journal} {Physical Review Letters}\ }\textbf {\bibinfo {volume} {124}}
  (\bibinfo {year} {2020}),\ 10.1103/physrevlett.124.116401}\BibitemShut
  {NoStop}%
\bibitem [{\citenamefont {Liao}\ \emph {et~al.}(2019)\citenamefont {Liao},
  \citenamefont {Li}, \citenamefont {Alavi},\ and\ \citenamefont
  {Gr{\"u}neis}}]{liao2019}%
  \BibitemOpen
  \bibfield  {author} {\bibinfo {author} {\bibfnamefont {K.}~\bibnamefont
  {Liao}}, \bibinfo {author} {\bibfnamefont {X.-Z.}\ \bibnamefont {Li}},
  \bibinfo {author} {\bibfnamefont {A.}~\bibnamefont {Alavi}}, \ and\ \bibinfo
  {author} {\bibfnamefont {A.}~\bibnamefont {Gr{\"u}neis}},\ }\href {\doibase
  10.1038/s41524-019-0243-7} {\bibfield  {journal} {\bibinfo  {journal} {npj
  Computational Materials}\ }\textbf {\bibinfo {volume} {5}},\ \bibinfo {pages}
  {110} (\bibinfo {year} {2019})}\BibitemShut {NoStop}%
\bibitem [{sup()}]{supp}%
  \BibitemOpen
  \href {http://dx.doi.org/} {\bibinfo  {journal} {See Supplemental Material}\
  }\BibitemShut {NoStop}%
\bibitem [{\citenamefont {Del~Ben}\ \emph {et~al.}(2015)\citenamefont
  {Del~Ben}, \citenamefont {Hutter},\ and\ \citenamefont
  {VandeVondele}}]{del2015}%
  \BibitemOpen
\bibfield  {journal} {  }\bibfield  {author} {\bibinfo {author} {\bibfnamefont
  {M.}~\bibnamefont {Del~Ben}}, \bibinfo {author} {\bibfnamefont
  {J.}~\bibnamefont {Hutter}}, \ and\ \bibinfo {author} {\bibfnamefont
  {J.}~\bibnamefont {VandeVondele}},\ }\href {\doibase 10.1063/1.4919238}
  {\bibfield  {journal} {\bibinfo  {journal} {The Journal of Chemical Physics}\
  }\textbf {\bibinfo {volume} {143}},\ \bibinfo {pages} {102803} (\bibinfo
  {year} {2015})}\BibitemShut {NoStop}%
\bibitem [{\citenamefont {Weigend}\ and\ \citenamefont
  {Häser}(1997)}]{weigend1997}%
  \BibitemOpen
  \bibfield  {author} {\bibinfo {author} {\bibfnamefont {F.}~\bibnamefont
  {Weigend}}\ and\ \bibinfo {author} {\bibfnamefont {M.}~\bibnamefont
  {Häser}},\ }\href {\doibase 10.1007/s002140050269} {\bibfield  {journal}
  {\bibinfo  {journal} {Theoretical Chemistry Accounts}\ }\textbf {\bibinfo
  {volume} {97}},\ \bibinfo {pages} {331} (\bibinfo {year} {1997})}\BibitemShut
  {NoStop}%
\bibitem [{\citenamefont {Rybkin}\ and\ \citenamefont
  {VandeVondele}(2016)}]{rybkin2016}%
  \BibitemOpen
  \bibfield  {author} {\bibinfo {author} {\bibfnamefont {V.~V.}\ \bibnamefont
  {Rybkin}}\ and\ \bibinfo {author} {\bibfnamefont {J.}~\bibnamefont
  {VandeVondele}},\ }\href {\doibase 10.1021/acs.jctc.6b00015} {\bibfield
  {journal} {\bibinfo  {journal} {Journal of Chemical Theory and Computation}\
  }\textbf {\bibinfo {volume} {12}},\ \bibinfo {pages} {2214} (\bibinfo {year}
  {2016})}\BibitemShut {NoStop}%
\bibitem [{\citenamefont {Kresse}\ and\ \citenamefont
  {Furthm\"uller}(1996)}]{kresse1996}%
  \BibitemOpen
  \bibfield  {author} {\bibinfo {author} {\bibfnamefont {G.}~\bibnamefont
  {Kresse}}\ and\ \bibinfo {author} {\bibfnamefont {J.}~\bibnamefont
  {Furthm\"uller}},\ }\href {http://link.aps.org/doi/10.1103/PhysRevB.54.11169}
  {\bibfield  {journal} {\bibinfo  {journal} {Phys. Rev. B}\ }\textbf {\bibinfo
  {volume} {54}},\ \bibinfo {pages} {11169} (\bibinfo {year} {1996})},\
  \bibinfo {note} {iSBN: 1098-0121}\BibitemShut {NoStop}%
\bibitem [{\citenamefont {Bl{\"{o}}chl}(1994)}]{blochl1994}%
  \BibitemOpen
  \bibfield  {author} {\bibinfo {author} {\bibfnamefont {P.~E.}\ \bibnamefont
  {Bl{\"{o}}chl}},\ }\href {\doibase 10.1103/PhysRevB.50.17953} {\bibfield
  {journal} {\bibinfo  {journal} {Phys. Rev. B}\ }\textbf {\bibinfo {volume}
  {50}},\ \bibinfo {pages} {17953} (\bibinfo {year} {1994})}\BibitemShut
  {NoStop}%
\bibitem [{\citenamefont {Hummel}\ \emph {et~al.}(2017)\citenamefont {Hummel},
  \citenamefont {Tsatsoulis},\ and\ \citenamefont {Grüneis}}]{hummel_2017}%
  \BibitemOpen
  \bibfield  {author} {\bibinfo {author} {\bibfnamefont {F.}~\bibnamefont
  {Hummel}}, \bibinfo {author} {\bibfnamefont {T.}~\bibnamefont {Tsatsoulis}},
  \ and\ \bibinfo {author} {\bibfnamefont {A.}~\bibnamefont {Grüneis}},\
  }\href {\doibase 10.1063/1.4977994} {\bibfield  {journal} {\bibinfo
  {journal} {J. Chem. Phys.}\ }\textbf {\bibinfo {volume} {146}},\ \bibinfo
  {pages} {124105} (\bibinfo {year} {2017})}\BibitemShut {NoStop}%
\bibitem [{\citenamefont {Solomonik}\ \emph {et~al.}(2014)\citenamefont
  {Solomonik}, \citenamefont {Matthews}, \citenamefont {Hammond}, \citenamefont
  {Stanton},\ and\ \citenamefont {Demmel}}]{solomonik2014}%
  \BibitemOpen
  \bibfield  {author} {\bibinfo {author} {\bibfnamefont {E.}~\bibnamefont
  {Solomonik}}, \bibinfo {author} {\bibfnamefont {D.}~\bibnamefont {Matthews}},
  \bibinfo {author} {\bibfnamefont {J.~R.}\ \bibnamefont {Hammond}}, \bibinfo
  {author} {\bibfnamefont {J.~F.}\ \bibnamefont {Stanton}}, \ and\ \bibinfo
  {author} {\bibfnamefont {J.}~\bibnamefont {Demmel}},\ }\href
  {https://www.sciencedirect.com/science/article/pii/S074373151400104X?via%3Dihub}
  {\bibfield  {journal} {\bibinfo  {journal} {Journal of Parallel and
  Distributed Computing}\ }\textbf {\bibinfo {volume} {74}},\ \bibinfo {pages}
  {3176} (\bibinfo {year} {2014})},\ \bibinfo {note} {publisher: Academic
  Press}\BibitemShut {NoStop}%
\bibitem [{\citenamefont {Gr\"uneis}\ \emph {et~al.}(2011)\citenamefont
  {Gr\"uneis}, \citenamefont {Booth}, \citenamefont {Marsman}, \citenamefont
  {Spencer}, \citenamefont {Alavi},\ and\ \citenamefont
  {Kresse}}]{gruneis2011}%
  \BibitemOpen
  \bibfield  {author} {\bibinfo {author} {\bibfnamefont {A.}~\bibnamefont
  {Gr\"uneis}}, \bibinfo {author} {\bibfnamefont {G.~H.}\ \bibnamefont
  {Booth}}, \bibinfo {author} {\bibfnamefont {M.}~\bibnamefont {Marsman}},
  \bibinfo {author} {\bibfnamefont {J.}~\bibnamefont {Spencer}}, \bibinfo
  {author} {\bibfnamefont {A.}~\bibnamefont {Alavi}}, \ and\ \bibinfo {author}
  {\bibfnamefont {G.}~\bibnamefont {Kresse}},\ }\href@noop {} {\bibfield
  {journal} {\bibinfo  {journal} {Journal of Chemical Theory and Computation}\
  }\textbf {\bibinfo {volume} {7}},\ \bibinfo {pages} {2780} (\bibinfo {year}
  {2011})}\BibitemShut {NoStop}%
\bibitem [{\citenamefont {Gr{\"u}neis}\ \emph {et~al.}(2010)\citenamefont
  {Gr{\"u}neis}, \citenamefont {Marsman},\ and\ \citenamefont
  {Kresse}}]{gruneis2010}%
  \BibitemOpen
  \bibfield  {author} {\bibinfo {author} {\bibfnamefont {A.}~\bibnamefont
  {Gr{\"u}neis}}, \bibinfo {author} {\bibfnamefont {M.}~\bibnamefont
  {Marsman}}, \ and\ \bibinfo {author} {\bibfnamefont {G.}~\bibnamefont
  {Kresse}},\ }\href {\doibase 10.1063/1.3466765} {\bibfield  {journal}
  {\bibinfo  {journal} {The Journal of Chemical Physics}\ }\textbf {\bibinfo
  {volume} {133}},\ \bibinfo {pages} {074107} (\bibinfo {year} {2010})},\
  \Eprint {http://arxiv.org/abs/https://doi.org/10.1063/1.3466765}
  {https://doi.org/10.1063/1.3466765} \BibitemShut {NoStop}%
\bibitem [{\citenamefont {Allen}\ and\ \citenamefont
  {Heine}(1976)}]{allen1976}%
  \BibitemOpen
  \bibfield  {author} {\bibinfo {author} {\bibfnamefont {P.~B.}\ \bibnamefont
  {Allen}}\ and\ \bibinfo {author} {\bibfnamefont {V.}~\bibnamefont {Heine}},\
  }\href {https://iopscience.iop.org/article/10.1088/0022-3719/9/12/013/meta}
  {\bibfield  {journal} {\bibinfo  {journal} {Journal of Physics C: Solid State
  Physics}\ }\textbf {\bibinfo {volume} {9}},\ \bibinfo {pages} {2305}
  (\bibinfo {year} {1976})}\BibitemShut {NoStop}%
\bibitem [{\citenamefont {Allen}\ and\ \citenamefont
  {Cardona}(1983)}]{allen1983}%
  \BibitemOpen
  \bibfield  {author} {\bibinfo {author} {\bibfnamefont {P.}~\bibnamefont
  {Allen}}\ and\ \bibinfo {author} {\bibfnamefont {M.}~\bibnamefont
  {Cardona}},\ }\href
  {https://journals.aps.org/prb/abstract/10.1103/PhysRevB.27.4760} {\bibfield
  {journal} {\bibinfo  {journal} {Physical Review B}\ }\textbf {\bibinfo
  {volume} {27}},\ \bibinfo {pages} {4760} (\bibinfo {year}
  {1983})}\BibitemShut {NoStop}%
\bibitem [{\citenamefont {Giannozzi}\ \emph {et~al.}(2009)\citenamefont
  {Giannozzi}, \citenamefont {Baroni}, \citenamefont {Bonini}, \citenamefont
  {Calandra}, \citenamefont {Car}, \citenamefont {Cavazzoni}, \citenamefont
  {Ceresoli}, \citenamefont {Chiarotti}, \citenamefont {Cococcioni},
  \citenamefont {Dabo},\ and\ \citenamefont {Dal~Corso}}]{giannozzi2009}%
  \BibitemOpen
  \bibfield  {author} {\bibinfo {author} {\bibfnamefont {P.}~\bibnamefont
  {Giannozzi}}, \bibinfo {author} {\bibfnamefont {S.}~\bibnamefont {Baroni}},
  \bibinfo {author} {\bibfnamefont {N.}~\bibnamefont {Bonini}}, \bibinfo
  {author} {\bibfnamefont {M.}~\bibnamefont {Calandra}}, \bibinfo {author}
  {\bibfnamefont {R.}~\bibnamefont {Car}}, \bibinfo {author} {\bibfnamefont
  {C.}~\bibnamefont {Cavazzoni}}, \bibinfo {author} {\bibfnamefont
  {D.}~\bibnamefont {Ceresoli}}, \bibinfo {author} {\bibfnamefont {G.~L.}\
  \bibnamefont {Chiarotti}}, \bibinfo {author} {\bibfnamefont {M.}~\bibnamefont
  {Cococcioni}}, \bibinfo {author} {\bibfnamefont {I.}~\bibnamefont {Dabo}}, \
  and\ \bibinfo {author} {\bibfnamefont {A.}~\bibnamefont {Dal~Corso}},\ }\href
  {https://iopscience.iop.org/article/10.1088/0953-8984/21/39/395502}
  {\bibfield  {journal} {\bibinfo  {journal} {J. Phys. Condens. Matter}\
  }\textbf {\bibinfo {volume} {21}},\ \bibinfo {pages} {395502} (\bibinfo
  {year} {2009})}\BibitemShut {NoStop}%
\bibitem [{\citenamefont {Marini}\ \emph {et~al.}(2009)\citenamefont {Marini},
  \citenamefont {Hogan}, \citenamefont {Gr{\"u}ning},\ and\ \citenamefont
  {Varsano}}]{marini2009}%
  \BibitemOpen
  \bibfield  {author} {\bibinfo {author} {\bibfnamefont {A.}~\bibnamefont
  {Marini}}, \bibinfo {author} {\bibfnamefont {C.}~\bibnamefont {Hogan}},
  \bibinfo {author} {\bibfnamefont {M.}~\bibnamefont {Gr{\"u}ning}}, \ and\
  \bibinfo {author} {\bibfnamefont {D.}~\bibnamefont {Varsano}},\ }\href
  {https://www.sciencedirect.com/science/article/pii/S0010465509000472}
  {\bibfield  {journal} {\bibinfo  {journal} {Comput. Phys. Commun.}\ }\textbf
  {\bibinfo {volume} {180}},\ \bibinfo {pages} {1392} (\bibinfo {year}
  {2009})}\BibitemShut {NoStop}%
\bibitem [{\citenamefont {Sangalli}\ \emph {et~al.}(2019)\citenamefont
  {Sangalli}, \citenamefont {Ferretti}, \citenamefont {Miranda}, \citenamefont
  {Attaccalite}, \citenamefont {Marri}, \citenamefont {Cannuccia},
  \citenamefont {Melo}, \citenamefont {Marsili}, \citenamefont {Paleari},
  \citenamefont {Marrazzo}, \citenamefont {Prandini}, \citenamefont {Bonfa},
  \citenamefont {Atambo}, \citenamefont {Affinito}, \citenamefont {Palummo},
  \citenamefont {Molina-Sanchez}, \citenamefont {Hogan}, \citenamefont
  {Gruning}, \citenamefont {Varsano},\ and\ \citenamefont
  {Marini}}]{sangalli2019}%
  \BibitemOpen
  \bibfield  {author} {\bibinfo {author} {\bibfnamefont {D.}~\bibnamefont
  {Sangalli}}, \bibinfo {author} {\bibfnamefont {A.}~\bibnamefont {Ferretti}},
  \bibinfo {author} {\bibfnamefont {H.}~\bibnamefont {Miranda}}, \bibinfo
  {author} {\bibfnamefont {C.}~\bibnamefont {Attaccalite}}, \bibinfo {author}
  {\bibfnamefont {I.}~\bibnamefont {Marri}}, \bibinfo {author} {\bibfnamefont
  {E.}~\bibnamefont {Cannuccia}}, \bibinfo {author} {\bibfnamefont
  {P.}~\bibnamefont {Melo}}, \bibinfo {author} {\bibfnamefont {M.}~\bibnamefont
  {Marsili}}, \bibinfo {author} {\bibfnamefont {F.}~\bibnamefont {Paleari}},
  \bibinfo {author} {\bibfnamefont {A.}~\bibnamefont {Marrazzo}}, \bibinfo
  {author} {\bibfnamefont {G.}~\bibnamefont {Prandini}}, \bibinfo {author}
  {\bibfnamefont {P.}~\bibnamefont {Bonfa}}, \bibinfo {author} {\bibfnamefont
  {M.~O.}\ \bibnamefont {Atambo}}, \bibinfo {author} {\bibfnamefont
  {F.}~\bibnamefont {Affinito}}, \bibinfo {author} {\bibfnamefont
  {M.}~\bibnamefont {Palummo}}, \bibinfo {author} {\bibfnamefont
  {A.}~\bibnamefont {Molina-Sanchez}}, \bibinfo {author} {\bibfnamefont
  {C.}~\bibnamefont {Hogan}}, \bibinfo {author} {\bibfnamefont
  {M.}~\bibnamefont {Gruning}}, \bibinfo {author} {\bibfnamefont
  {D.}~\bibnamefont {Varsano}}, \ and\ \bibinfo {author} {\bibfnamefont
  {A.}~\bibnamefont {Marini}},\ }\href {<Go to ISI>://WOS:000469804900001}
  {\bibfield  {journal} {\bibinfo  {journal} {J. Phys.: Condens. Matter}\
  }\textbf {\bibinfo {volume} {31}},\ \bibinfo {pages} {325902} (\bibinfo
  {year} {2019})}\BibitemShut {NoStop}%
\bibitem [{\citenamefont {Lee}\ \emph {et~al.}(2010)\citenamefont {Lee},
  \citenamefont {Murray}, \citenamefont {Kong}, \citenamefont {Lundqvist},\
  and\ \citenamefont {Langreth}}]{lee2010}%
  \BibitemOpen
  \bibfield  {author} {\bibinfo {author} {\bibfnamefont {K.}~\bibnamefont
  {Lee}}, \bibinfo {author} {\bibfnamefont {E.~D.}\ \bibnamefont {Murray}},
  \bibinfo {author} {\bibfnamefont {L.}~\bibnamefont {Kong}}, \bibinfo {author}
  {\bibfnamefont {B.~I.}\ \bibnamefont {Lundqvist}}, \ and\ \bibinfo {author}
  {\bibfnamefont {D.~C.}\ \bibnamefont {Langreth}},\ }\href {\doibase
  10.1103/PhysRevB.82.081101} {\bibfield  {journal} {\bibinfo  {journal} {Phys.
  Rev. B}\ }\textbf {\bibinfo {volume} {82}},\ \bibinfo {pages} {081101}
  (\bibinfo {year} {2010})}\BibitemShut {NoStop}%
\bibitem [{\citenamefont {Gruber}\ \emph {et~al.}(2018)\citenamefont {Gruber},
  \citenamefont {Liao}, \citenamefont {Tsatsoulis}, \citenamefont {Hummel},\
  and\ \citenamefont {Gr\"uneis}}]{gruber2018}%
  \BibitemOpen
  \bibfield  {author} {\bibinfo {author} {\bibfnamefont {T.}~\bibnamefont
  {Gruber}}, \bibinfo {author} {\bibfnamefont {K.}~\bibnamefont {Liao}},
  \bibinfo {author} {\bibfnamefont {T.}~\bibnamefont {Tsatsoulis}}, \bibinfo
  {author} {\bibfnamefont {F.}~\bibnamefont {Hummel}}, \ and\ \bibinfo {author}
  {\bibfnamefont {A.}~\bibnamefont {Gr\"uneis}},\ }\href
  {https://link.aps.org/doi/10.1103/PhysRevX.8.021043} {\bibfield  {journal}
  {\bibinfo  {journal} {Physical Review X}\ }\textbf {\bibinfo {volume} {8}},\
  \bibinfo {pages} {21043} (\bibinfo {year} {2018})},\ \bibinfo {note}
  {publisher: American Physical Society}\BibitemShut {NoStop}%
\bibitem [{\citenamefont {Shen}\ \emph {et~al.}(2020)\citenamefont {Shen},
  \citenamefont {Zhang}, \citenamefont {Shang}, \citenamefont {Zhang},
  \citenamefont {Wang}, \citenamefont {Wang}, \citenamefont {Jiang},\ and\
  \citenamefont {Li}}]{shen2020}%
  \BibitemOpen
  \bibfield  {author} {\bibinfo {author} {\bibfnamefont {T.}~\bibnamefont
  {Shen}}, \bibinfo {author} {\bibfnamefont {X.-W.}\ \bibnamefont {Zhang}},
  \bibinfo {author} {\bibfnamefont {H.}~\bibnamefont {Shang}}, \bibinfo
  {author} {\bibfnamefont {M.-Y.}\ \bibnamefont {Zhang}}, \bibinfo {author}
  {\bibfnamefont {X.}~\bibnamefont {Wang}}, \bibinfo {author} {\bibfnamefont
  {E.-G.}\ \bibnamefont {Wang}}, \bibinfo {author} {\bibfnamefont
  {H.}~\bibnamefont {Jiang}}, \ and\ \bibinfo {author} {\bibfnamefont {X.-Z.}\
  \bibnamefont {Li}},\ }\href
  {http://journals.aps.org/prb/abstract/10.1103/PhysRevB.102.045117} {\bibfield
   {journal} {\bibinfo  {journal} {Phys. Rev. B}\ }\textbf {\bibinfo {volume}
  {102}},\ \bibinfo {pages} {045117} (\bibinfo {year} {2020})}\BibitemShut
  {NoStop}%
\bibitem [{\citenamefont {Silvera}\ and\ \citenamefont
  {Dias}(2018)}]{silvera2018}%
  \BibitemOpen
  \bibfield  {author} {\bibinfo {author} {\bibfnamefont {I.~F.}\ \bibnamefont
  {Silvera}}\ and\ \bibinfo {author} {\bibfnamefont {R.}~\bibnamefont {Dias}},\
  }\href {\doibase 10.1088/1361-648X/aac401} {\bibfield  {journal} {\bibinfo
  {journal} {Journal of Physics: Condensed Matter}\ }\textbf {\bibinfo {volume}
  {30}},\ \bibinfo {pages} {254003} (\bibinfo {year} {2018})}\BibitemShut
  {NoStop}%
\bibitem [{\citenamefont {Loubeyre}\ \emph {et~al.}(2017)\citenamefont
  {Loubeyre}, \citenamefont {Occelli},\ and\ \citenamefont
  {Dumas}}]{loubeyre2017}%
  \BibitemOpen
  \bibfield  {author} {\bibinfo {author} {\bibfnamefont {P.}~\bibnamefont
  {Loubeyre}}, \bibinfo {author} {\bibfnamefont {F.}~\bibnamefont {Occelli}}, \
  and\ \bibinfo {author} {\bibfnamefont {P.}~\bibnamefont {Dumas}},\ }\href
  {http://arxiv.org/abs/1702.07192} {\bibfield  {journal} {\bibinfo  {journal}
  {arXiv:1702.07192 [cond-mat]}\ } (\bibinfo {year} {2017})},\ \bibinfo {note}
  {arXiv: 1702.07192}\BibitemShut {NoStop}%
\end{thebibliography}%



\end{document}